\documentstyle[aps,pre,floats]{revtex}

\begin{document}

\twocolumn[\hsize\textwidth\columnwidth\hsize\csname@twocolumnfalse%
\endcsname

\title{A scaling theory of 3D spinodal turbulence}
\author{V. M. Kendon\footnote{
Current address: Dept. of Physics and Applied Physics,
University of Strathclyde, Glasgow, G1 1XQ, United Kingdom}
}
\address{Department of Physics and Astronomy, 
JCMB King's Buildings, University of Edinburgh, Mayfield Road,\\
Edinburgh EH9 3JZ, United Kingdom}
\maketitle
\vspace{-0.4cm}
\center{\small{(Received 18 February 2000)}}
\begin{abstract}
A new scaling theory for spinodal decomposition in the inertial hydrodynamic
regime is presented. The scaling involves three relevant length scales,
the domain size, the Taylor microscale and the Kolmogorov dissipation scale. 
This allows for the presence of an inertial ``energy cascade'', familiar
from theories of turbulence, and improves on earlier scaling treatments
based on a single length: these, it is shown, cannot be reconciled with
energy conservation. The new theory reconciles the $t^{2/3}$ scaling of the
domain size, predicted by simple scaling, with the physical expectation 
of a saturating Reynolds number at late times.
\hfill
{\rule[0ex]{4ex}{0ex}}
\end{abstract}
\bigskip
\footnotesize{\hspace{2cm}PACS numbers: 64.75+g}\hfill
\bigskip
]


A binary fluid mixture will undergo phase separation, if the two fluids are
mutually repulsive, below some critical temperature.  Presented here is a
theoretical study of the dynamics of spinodal decomposition in three
dimensions, in a 50/50 mixture where the two fluids are incompressible and have
the same shear viscosity, $\eta$, and density, $\rho$.
Starting from a completely mixed state quenched to far below the critical
temperature, the initial separation is dominated by diffusion
until well-defined interfaces form between two
interlocking domains of single fluid regions.
The ensuing late-stage coarsening is then driven by the 
interfacial tension, resisted by viscosity as the bulk fluid
flows so as to flatten the interfaces and enlarge the domain size.
The coarsening can be followed through the average
domain size, $L$, most commonly measured by
the inverse first moment of the spherically
averaged structure factor, $S(k)$ of the difference in the concentrations of
the two fluids,
$L = 2\pi  \int\!S(k)\,{\mathrm d}k / \int\!k S(k)\,{\mathrm d}k$.
Using a simple (single length scale) scaling theory, Siggia \cite{siggia79a}
predicted that the domain size first shows linear growth,
$L\sim t$, in the viscous hydrodynamic regime.
Furukawa \cite{furukawa85a} extended this, predicting 
a crossover to a slower growth rate of $t^{2/3}$
in the inertial hydrodynamic regime.
These growth rates have been observed in numerical simulation,
see, e.g., \cite{appert95a,laradji96a,bastea97a,jury98a,kendon99a}.
Linear growth has been observed experimentally \cite{foot:expt}.

In what follows, I show that the simple scaling theory is
inconsistent with energy conservation in the inertial regime; a minimal
alternative is presented, based on three relevant length scales, which
allows both force balance and conservation of energy to be maintained at
late times.  The new scaling theory recovers the $L \sim t^{2/3}$
scaling, but gives altered scalings for other quantities (such as velocity
gradients). This allows the physical requirement,
recently emphasized by Grant and Elder \cite{grant99a},
that the fluid Reynolds number should not diverge in the long time limit,
to be reconciled with the $t^{2/3}$ scaling.

The scaling approach developed here ignores the anomalous scaling
corrections of modern turbulence theory \cite{turb}.  The analysis of
spinodal decomposition presented below is thus an approximate one,
but as legitimate as the Kolmogorov theory of homogeneous turbulence.
As such, it is likely to be a useful tool for interpreting simulations and
experiments on the inertial hydrodynamic regime of spinodal decomposition,
and may be the best description available until the (simpler) problem of
homogeneous turbulence is fully solved.



The system can be described by the isothermal incompressible
Navier-Stokes equation (NSE),
\begin{equation}
\rho\frac{\partial{\mathbf v}}{\partial t} + \rho({\mathbf v}\cdot\nabla){\mathbf v}
 = \eta\nabla^2{\mathbf v} -\nabla\cdot{\mathcal P},
\label{eq:nse}
\end{equation}
where ${\mathbf v}$ is the fluid velocity.  (This can be represented by a
single variable regardless of fluid composition since the two fluids have
identical properties.)
Included in the pressure tensor, ${\mathcal P}$, is the interfacial stress
which comes from capillary forces; the excess Laplace pressure
in the curved interface drives the bulk fluid away from regions
of tight curvature which eventually collapse into narrow ``necks''
that break, leading to further enlargment of the remaining bulk domains.
Using simple scaling arguments \cite{siggia79a,furukawa85a,bray94a},
the interfacial force density, $\nabla\cdot{\mathcal P}$,
can be approximated by $\sigma/L^2$,
where $\sigma$ is the interfacial tension between the two fluids.
The interface is assumed to remain locally smooth and completely percolated
throughout the phase separation process, with constant $\sigma$.
This is reasonable provided diffusion is rapid on the scale, $\xi$, of the
interfacial width (to maintain local equilibrium on the time scale of
the interfacial motion) while being negligible over scales of the
order of the domain size so it does not contribute to the domain growth rate.
This requires care in simulation work \cite{kendon99a}, but in real fluids,
$L \gg \xi$ can easily be arranged.

The linear scaling is readily obtained from Eq. (\ref{eq:nse}) by
neglecting the inertial terms on the l.h.s. and equating the
viscous term to the interfacial one, using ${\mathbf v} \sim \dot{L}$
and $\nabla \sim 1/L$.
When the inertial terms are no longer negligible,
reversing this argument and equating the
interfacial force to the inertial terms, produces $L \sim t^{2/3}$.
This assumes that there is only one
relevant macroscopic length scale in the system.  (Clearly there is
also a microscopic length scale in the width of the interface, but it will
be assumed, as usual, that this does not affect the macrosopic growth
\cite{foot:nse}.)

A Reynolds number can be defined \cite{furukawa85a}
by ${\mathrm Re}_L = \rho L \dot{L}/\eta$.
Clearly for $L \sim t^{2/3}$, ${\mathrm Re}_L \sim t^{1/3}$, and thus
${\mathrm Re}_L$ is predicted to grow without bound in the inertial regime.
Recently, Grant and Elder \cite{grant99a} suggested that this
unbounded growth of the Reynolds number is unphysical,
and there should be a further crossover of
the domain growth rate to $t^{1/2}$ (or slower)
in order that ${\mathrm Re}_L$ remains finite.
However, the first simulation studies to reliably reach the inertial regime
\cite{kendon99a} did not find evidence of any final growth rate
slower than $t^{2/3}$.
Below it is argued that ${\mathrm Re}_L$ does not quantify the inertial
effects which occur in the bulk fluid, which are instead
properly characterised by the Reynolds number, ${\mathrm Re}_{\mathbf v} =
\rho|{\mathbf v}\cdot\nabla{\mathbf v}|/(\eta|\nabla^2{\mathbf v}|)$, 
that is, the ratio of the nonlinear to the viscous terms in the NSE,
and that while ${\mathrm Re}_L$ diverges, ${\mathrm Re}_{\mathbf v}$
does not.
It is easily seen that ${\mathrm Re}_{\mathbf v}$ must remain finite.
Infinite ${\mathrm Re}_{\mathbf v}$ implies 
an infinite energy density, while the
fluid mixture starts with a finite energy density from the free energy
difference between the mixed and separated states.



I start from the energy equation for the system, which (ignoring heat flow)
may be written,
\begin{eqnarray}
\frac{\partial}{\partial t}(\rho v^2/2) &+& {\mathbf v}\cdot\nabla(\rho v^2/2) =
\nonumber \\
&-&  \eta\left(\nabla{\mathbf v}\right)^2 + {\mathbf v}\cdot\nabla\cdot{\mathcal P}
+ (\eta/\rho)\nabla^2(\rho v^2/2),
\label{eq:l_ebal}
\end{eqnarray}
where $\left(\nabla{\mathbf v}\right)^2\equiv \left(\nabla{\mathbf v}\right):
\left(\nabla{\mathbf v}\right)$.
Since we are not concerned with the local convective or diffusive energy
flows, it is more convenient to average over the whole system and
write the global energy balance per unit volume as,
\begin{equation}
\frac{{\mathrm d}}{{\mathrm d}t}\left\langle\rho v^2/2\right\rangle
= - \eta\left\langle\left(\nabla{\mathbf v}\right)^2\right\rangle
  + \varepsilon_{{\mathrm in}},
\label{eq:ebal}
\end{equation}
where $\varepsilon_{{\mathrm in}}$ is the rate of energy transfer
to the fluid from the interface, which can be approximated by
$\sigma\dot{L}/L^2$ (force $\cdot$ velocity).

I now show that, in the inertial hydrodynamic regime,
the simple scaling theory is inconsistent with
energy conservation, by considering the behavior of the global energy balance 
under the scaling predicted for this regime \cite{furukawa85a}.
Applying the simple scaling to each term in Eq. (\ref{eq:ebal}),
and replacing $L$ by $t^{2/3}$ gives,
\begin{equation}
-\rho t^{-5/3} \sim -\eta t^{-2} + \sigma t^{-5/3}.
\end{equation}
The dissipation term ($-\eta t^{-2}$) clearly becomes negligible compared
to the other two terms, as originally assumed.
However, the kinetic energy in the fluid
and the energy stored in the interface are \textit{both decreasing} over time;
thus energy conservation cannot possibly be maintained except by including the
dissipation term in the energy balance equation.
Likewise, the viscous term should never be neglected in the NSE
because it involves the highest order in derivatives and is thus a
singular perturbation: however small $\eta$ is, the
asymptotic physics is radically altered from that with $\eta=0$.



I now allow for more general scaling behavior in
the NSE and global energy balance equations
by introducing two new lengths, $L_{\nabla}$ and $L_{\nabla^2}$,
with associated scaling exponents $\alpha'$ and $\alpha''$, for the
velocity first and second derivatives respectively, and
allowing the velocity to scale as $t^{\beta}$.
Terms associated with the interface are still assumed to have
scaling determined by the domain size.
The scaling quantities are defined as follows,
\begin{eqnarray}
\mbox{domain size:\hspace{2ex}} && L \sim t^{\alpha} \nonumber \\
\mbox{fluid velocity:\hspace{2ex}} && {\mathbf v} \sim t^{\beta} \nonumber \\
\mbox{velocity first derivative:\hspace{2ex}} &&
	\nabla{\mathbf v} \sim {\mathbf v}/L_{\nabla} \sim t^{\beta-\alpha'}
	\nonumber \\
\mbox{velocity second derivative:\hspace{2ex}} &&
      \nabla^2 {\mathbf v} \sim {\mathbf v}/L^2_{\nabla^2} \sim
	t^{\beta-2\alpha''}. \nonumber
\end{eqnarray}
This supposes that the interface remains smooth, but allows nontrivial
structure in the fluid velocity at smaller scales.
Using these scalings to write the NSE, Eq. (\ref{eq:nse}),
and energy balance equation, Eq. (\ref{eq:ebal}), as powers of $t$ gives,
respectively,
\begin{eqnarray}
\mbox{NSE: } && \rho\beta t^{\beta-1} + \rho t^{2\beta-\alpha'}
          \sim \,\,\eta t^{\beta-2\alpha''} + \sigma t^{-2\alpha}
\label{eq:b_scale1} \\
\mbox{energy: } && \rho\beta t^{2\beta-1} \mbox{\hspace{1.4cm}}
          \sim -\eta t^{2\beta-2\alpha'} + \sigma t^{-\alpha-1},
\label{eq:b_scale2}
\end{eqnarray}
where the prefactors have been left in to facilitate
identification of the terms.
This more general scaling ignores the anomalous scaling corrections
of modern turbulence theory \cite{turb}, but is sufficient to make
progress on the problem of spinodal decomposition.
The local energy equation, Eq. (\ref{eq:l_ebal}), becomes,
\begin{eqnarray}
\rho\beta t^{2\beta-1} &+& \rho t^{3\beta-\alpha'} \sim \nonumber \\
          &-& \eta t^{2\beta-2\alpha'} + \sigma t^{-\alpha-1}
               +(\eta/\rho) t^{2\beta-2\alpha''}.
\label{eq:b_lscale}
\end{eqnarray}
Solutions for $\alpha$, $\alpha'$, $\alpha''$ and $\beta$ in
Eqs. (\ref{eq:b_scale1}) and (\ref{eq:b_scale2}) will later be checked in
Eq. (\ref{eq:b_lscale}) to confirm that there are no
discrepancies predicted for the behavior of the local energy flows within this
three-length-scaling analysis.

As already pointed out, dissipation must remain significant, so
we look for a three-way balance between the terms in
the energy balance equation, Eq. (\ref{eq:b_scale2}), giving for the exponents,
\begin{equation}
2\beta-1 = 2\beta-2\alpha' = -\alpha-1.
\end{equation}
This gives $\alpha'=1/2$, and $\beta=-\alpha/2$.  Substituting these back
into the NSE, Eq. (\ref{eq:b_scale1}), gives,
\begin{equation}
\rho t^{-\alpha/2-1} + \rho t^{-\alpha-1/2} \sim
\eta t^{-\alpha/2-2\alpha''} + \sigma t^{-2\alpha}.
\end{equation}
There is no solution to this with all four terms having the same
exponent; solutions can instead be found by balancing the terms off in pairs.
Numbering the terms 1--4 from left to right, the pairing that gives the
inertial regime scaling is
term 1 with term 4, and term 2 with term 3, giving $\alpha=2/3$, and
$\alpha''=5/12$, with $\beta=-\alpha/2=-1/3$.
The four terms in the NSE thus scale as follows,
\begin{equation}
\rho t^{-4/3} + \rho t^{-7/6} \sim \eta t^{-7/6} + \sigma t^{-4/3}.
\end{equation}
The physical interpretation of this 
is that the moving interface gives rise to large scale velocity motion
via $\rho\partial{\mathbf v}/\partial t$.  The nonlinear term,
$\rho{\mathbf v}\cdot\nabla{\mathbf v}$, then transfers the energy from large
length scales to small length scales where it is finally removed
by dissipation.  This is the familiar ``energy cascade'' of turbulence theory.
Note that scaling arguments only predict that the paired terms
balance approximately, so the nonlinear term,
$\rho{\mathbf v}\cdot\nabla {\mathbf v}$, is, in general,
larger than the viscous term, $\eta\nabla^2{\mathbf v}$,
due to the energy ``in transit'' from large to small scales,
contained in a series of eddies of decreasing size.
The transverse components of $\rho\partial{\mathbf v}/\partial t$ are thus
also larger than required to balance the interfacial force, to account for
the rotational motion from $\rho{\mathbf v}\cdot\nabla {\mathbf v}$.
The length scales associated with $\nabla$ and $\nabla^2$ both grow more
slowly than $L$, with $L_{\nabla}\sim t^{1/2}$,
and $L_{\nabla^2}\sim t^{5/12}$, so there is an increasing separation of
length scales within the system.  The dissipation is thus decoupled
from the interfacial energy input, and no longer affects the
domain growth rate.
The Reynolds number, defined as the ratio of the nonlinear to viscous terms,
${\mathrm Re}_{\mathbf v} = \rho|{\mathbf v}\cdot\nabla {\mathbf v}|/
(\eta|\nabla^2 {\mathbf v}|)$,
remains finite (satisfying the physical demand of Grant and Elder
\cite{grant99a}),
while the domain size grows as $t^{2/3}$ (contrary to their deduction that
$\alpha \le \frac{1}{2}$).
With this scaling, the local energy equation, Eq. (\ref{eq:b_lscale}),
becomes,
\begin{equation}
-\frac{\rho}{3}t^{-5/3} + \rho t^{-3/2} \sim
-\eta t^{-5/3} + \sigma t^{-5/3} + \frac{\eta}{\rho} t^{-3/2}.
\end{equation}
The local convective and diffusive terms are dominant, and balance each other,
representing the energy moved around by the turbulent fluid flow.

These results are summarised in Table \ref{table:predict}, alongside the
predictions of the simple scaling theory, for comparison.
(Results for the viscous hydrodynamic regime are also shown.)
In particular, notice that in the new scaling theory for the inertial regime,
the lengths $L_{\nabla}$ and $L_{\nabla^2}$ have
the same scaling as $\lambda =(5\eta\langle v^2\rangle/\varepsilon)^{1/2}$,
the Taylor microscale, and $\lambda_d=2\pi(\eta^3/\rho^3\varepsilon)^{1/4}$,
the Kolmogorov dissipation scale respectively.
The Taylor microscale characterises the length scales in a turbulent fluid
at which dissipation becomes significant, while the Kolmogorov dissipation
scale marks the small-scale end of the dissipation range \cite{monin75a}.
Thus, although these new scaling results have been obtained without
specific input from turbulence theory, the extra length scales
for the velocity derivatives turn out to (within prefactors) coincide with
key characteristic quantities in turbulence phenomenology,
providing strong support for the new theory.



It is easily shown that
the familiar linear scaling of the viscous regime \cite{siggia79a}
is consistent with energy conservation by substituting
$\alpha=\alpha'=\alpha''=1$ and $\beta=0$ into
Eqs. (\ref{eq:b_scale1}) and (\ref{eq:b_scale2}), giving,
\begin{eqnarray}
\rho t^{-1} &\sim& \eta t^{-2} + \sigma t^{-2},
\label{eq:scale_nse0} \\
\eta t^{-2} &\sim& \sigma t^{-2},
\label{eq:scale_e0}
\end{eqnarray}
respectively.
In the energy equation, Eq. (\ref{eq:scale_e0}), there is a simple balance
between energy input and dissipation, but in the NSE, Eq. (\ref{eq:scale_nse0}),
the nonlinear term on the l.h.s.,
which was assumed to be negligible in the original simple scaling argument,
appears to be decaying more slowly ($t^{-1}$) than the r.h.s. terms ($t^{-2})$.
Recalling that the viscous regime is not the long time asymptotic regime,
and only expected to hold for times earlier than some 
crossover time (before which the nonlinear term will be smaller
than the other two terms), this apparent difficulty is eliminated.



There are a few further potential solutions to the exponents in Eqs.
(\ref{eq:b_scale1}) and (\ref{eq:b_scale2}),
which I will now discuss briefly.  All can be eliminated on physical
grounds.  In the inertial regime, where the NSE terms are balanced in
pairs, the other possible pairings of terms must checked.
It is not possible to balance term 1 with term 2 and term 3
with term 4 because the nonlinear term (term 2)
is a force perpendicular to ${\mathbf v}$ (this is obvious when it is
written in the alternative form, $-\rho{\mathbf v}\times\nabla\times{\mathbf v}$)
and therefore it cannot change the magnitude of ${\mathbf v}$, impying
$\beta=0$.  However, term 1 is proportional to $\beta$, so is only
non-zero if $\beta\ne 0$.
Matching term 1 with term 3, and term 2 with term 4
gives a solution for the exponents of,
$\alpha=1/2$, $\alpha'=1/2$, $\alpha''=1/2$ and $\beta=-1/4$.
Physically, this solution has just one length scale in the system,
but the velocity is decoupled from the interface,
${\mathbf v}\sim t^{-1/4}$, while $\dot{L}(t)\sim t^{-1/2}$.
In the local energy equation, Eq. (\ref{eq:b_lscale}), this 
solution becomes,
\begin{equation}
-\frac{\rho}{4}t^{-3/2} + \rho t^{-5/4} \sim
-\eta t^{-3/2} + \sigma t^{-3/2} + \frac{\eta}{\rho} t^{-3/2}.
\end{equation}
The convective term, $\rho t^{-5/4}$, will eventually come to
dominate over all the other terms, thus it cannot represent an
asymptotic solution for late times.
There is one further solution for $\beta=0$,
obtained by solving Eqs. (\ref{eq:b_scale1}) and (\ref{eq:b_scale2})
for the remaining exponents, $\alpha$, $\alpha'$, and $\alpha''$, giving,
$\alpha=1/3$, $\alpha'=2/3$ and $\alpha''=1/3$.
The length scale for velocity gradients is
related to the domain size by $L_{\nabla}\sim L^2(t)$, suggesting that
the nonlinear term is mixing on scales larger than the domain size.
Physically, this could correspond to a ``turbulent remixing'' regime as
suggested by Grant and Elder \cite{grant99a}.
However, the resulting breakup of the interface is liable to invalidate
the assumptions made in deriving the scaling approximation for the
interfacial force.  Thus, although this appears to be a consistent solution
to the NSE and energy balance equations, it seems unlikely that
the system could, in fact, ever achieve such a scaling.



In summary, I have obtained consistent scaling behavior
for domain growth in the spinodal decomposition of a symmetric binary fluid
mixture by including two extra macroscopic length scales
for the velocity derivatives to enable energy balance to be satisfied.
The velocity itself is still found to scale as $\dot{L}(T)$.
In a system restricted to derivatives up to second order
(the NSE is, itself, an approximation based on separation of macroscopic
and molecular length and time scales), one length scale per derivative would 
seem to be a reasonable maximum for the purpose of simple scaling arguments,
and it does not seem to be possible to satisfy the NSE and
energy equations using fewer.

With this new scaling, the inertial regime scaling of $L\sim t^{2/3}$ is
maintained while the fluid Reynolds number, ${\mathrm Re}_{\mathbf v}$,
measured as the actual
ratio of the nonlinear to the viscous terms in the NSE, remains finite.
The Reynolds number obtained from the domain size, ${\mathrm Re}_L$,
is not a good estimate of ${\mathrm Re}_{\mathbf v}$ in the inertial regime
so it is not
physically significant that ${\mathrm Re}_L$ continues to grow without bound.
The key point for the scaling behavior of $L$ is that the driving force
from the interface balances against the acceleration term alone,
so is decoupled from what happens in the remainder of the fluid motion.
The nonlinear and viscous terms are free to find their own 
balance independent of the driving force, provided that they
do, eventually, remove energy from the system.

There is, as yet, no simulation or experimental work that tests this new
scaling theory.  The simulation work reported in
\cite{kendon99a,kendon99b}, while showing hints of possible different scaling
for the NSE terms, was not able to
probe far enough into the inertial regime to provide a significant test.
Using $L_0=\eta^2/(\rho\sigma)$ and $t_0=\eta^3/(\rho\sigma^2)$
to provide non-dimensional length ($L/L_0$) and time ($t/t_0$), 
Ref. \cite{kendon99a} suggests that $L/L_0 > 10^6$, $t/t_0 > 10^9$ will
be required, at least an order of magnitude beyond present simulation
capabilities.
Experimental work, though more feasible, will not be easy to perform.
Using values for water, $\eta = 10^{-3}$ Kg m$^{-1}$ s$^{-1}$,
and for water-paraffin, $\sigma = 2.4$ $10^{-2}$ Kg s$^{-2}$,
with the projection from simulation work of $L/L_0 > 10^6$
to estimate the interfacial force ($\sigma/L^2$),
the density matching of the two mutually repulsive fluids
must be better than one part in $10^4$ for terrestrial gravitational effects
to remain small in the required regime ($L\sim 4$ cm).
The micro-gravity environment of the space shuttle would ease this constraint
\cite{beysens97a}.

I would like to thank Mike Cates and Ignacio Pagona\-barraga for valuable
discussions and a careful reading of the manuscript,
and EPSRC for a PhD studentship.  This work was
funded in part by EPSRC GR/M56234.




\begin{onecolumn}
\begin{table}
    \begin{minipage}{\textwidth}
        \renewcommand{\arraystretch}{1.1}
        \begin{center}
        \begin{tabular}{llrrr}
            \multicolumn{2}{c}{\textit{quantity}} & \textit{viscous} &
		\multicolumn{2}{c}{\textit{inertial regime}}   \\
            &&\textit{regime} & \textit{simple scaling} &
		\textit{new scaling} \\
            \hline
            domain size &$L$ & 1 & 2/3 & 2/3 \\
            length for fluid velocity first derivative
		&$L_{\nabla}$ & 1 & 2/3 & \textbf{1/2} \\
            length for fluid velocity second derivative&
		$L_{\nabla^2}$ & 1 & 2/3 & \textbf{5/12} \\
            fluid velocity &${\mathbf v}$ & 0 & -1/3 & -1/3 \\
            \hline
            &$\rho\partial{\mathbf v}/\partial t$ & $=0$ & -4/3 & -4/3 \\
            NSE
	    	&$\rho{\mathbf v}\cdot\nabla{\mathbf v}$ & $=0/{\mathbf -1}$ & -4/3 & \textbf{-7/6} \\
            terms
	    	&$\eta\nabla^2{\mathbf v}$ & -2 & -5/3 & \textbf{-7/6} \\
            &$\sigma/L^2$ & -2 & -4/3 & -4/3 \\
            \hline
            Reynolds number from interface
		&${\mathrm Re}_L=(\rho/\eta)L\dot{L}$ & 1 & 1/3 & 1/3 \\
            Reynolds number in fluid
	      &${\mathrm Re}_{\mathbf v} = \rho{\mathbf v}\cdot\nabla{\mathbf v}/
		\eta\nabla^2{\mathbf v}$ & $=0$ & 1/3 & \textbf{0} \\
            dissipation rate &$\varepsilon=\eta(\nabla{\mathbf v})^2$
		& -2 & -2 & \textbf{-5/3} \\
            Taylor microscale
		&$\lambda=(5\eta\langle v^2\rangle/\varepsilon)^{1/2}$
		& 1 & 2/3 &
 		\textbf{1/2} \\
            Kolmogorov dissipation scale
		&$\lambda_d=2\pi(\eta^3/\rho^3\varepsilon)^{1/4}$ & 1/2 & 1/2 &
		\textbf{5/12} \\
        \end{tabular}
        \end{center}
        \caption{\textit{Summary of predicted scaling exponents for
        the viscous and inertial regimes.  The new theory has the same
	predictions for the viscous regime as the simple theory, apart from
	the NSE term $\rho{\mathbf v}\cdot\nabla{\mathbf v}$.  Entries
	are powers of time, $t$; an entry of $0$ indicates the quantity is
        constant, while an entry of $=0$ indicates the quantity is assumed
        to be zero in the viscous approximation.
        Bold entries indicate new scaling predictions that differ from
        the simple theory.}}
        \protect\label{table:predict}
    \end{minipage}
\end{table}
\end{onecolumn}

\end{document}